\documentclass[final,twocolumn]{IEEEtran}

\usepackage{amsmath}

\usepackage{graphicx}
\usepackage{epsfig}
\usepackage{epstopdf}
\usepackage{subfigure,color}
\usepackage{dcolumn}
\usepackage{bm}

\usepackage{cite}

\def\e{\begin{equation}}
\def\f{\end{equation}} 

\providecommand*{\mrm}[1]{\mathrm{#1}}

\renewcommand{\Re}{\ensuremath{\mrm{Re}}}	

\def\XXint#1#2#3{{\setbox0=\hbox{$#1{#2#3}{\int}$}
     \vcenter{\hbox{$#2#3$}}\kern-.5\wd0}}

\begin{document}
	

\title{Physical meaning of the dipole radiation resistance in lossless and lossy media}
\author{M.~S.~Mirmoosa, \IEEEmembership{Member, IEEE}, S.~Nordebo, and S.~A.~Tretyakov, \IEEEmembership{Fellow, IEEE}\thanks{This work has been partly supported by the Swedish Foundation for Strategic Research (SSF) under the programme Applied Mathematics and the project Complex Analysis and Convex Optimization for EM Design.}\thanks{M.~S.~Mirmoosa and S.~A.~Tretyakov are with the Department of Electronics and Nanoengineering, Aalto University, P.O.~Box 15500, FI-00076 Aalto, Finland (email: mohammad.mirmoosa@aalto.fi and sergei.tretyakov@aalto.fi).}\thanks{S.~Nordebo is with the Department of Physics and Electrical Engineering, Linn{\ae}us University, 351~95~V{\"a}xj{\"o}, Sweden (email: sven.nordebo@lnu.se).}}

\maketitle

\begin{abstract}

In this tutorial, we discuss the radiation from a Hertzian dipole into uniform isotropic lossy media of infinite extent. If the medium is lossless, the radiated power propagates to infinity, and the apparent dissipation is measured by the radiation resistance of the dipole. If the medium is lossy, the power exponentially decays, and the physical meaning of radiation resistance needs clarification. Here, we present explicit calculations of the power absorbed in the infinite lossy host space and discuss the limit of zero losses. We show that the input impedance of dipole antennas contains a radiation-resistance contribution which does not depend on the imaginary part of the refractive index. This fact means that the power delivered by dipole antennas to surrounding space always contains a contribution from far fields unless the real part of the refractive index is zero. Based on this understanding, we discuss the fundamental limitations of power coupling between two antennas and possibilities of removing the limit imposed by radiation damping.

\end{abstract}


\section{Introduction}

The radiation resistance of a dipole antenna is a simple classical concept which is used to model the power radiated into surrounding infinite space, as explained in many antenna textbooks, e.g. \cite{balanis}. However, its physical meaning is not always easy to grasp. Indeed, if the surrounding space is lossless, the energy  cannot be dissipated. Yet, if the space around the antenna is infinite, the radiation resistance is nonzero, which is tantamount to absorption of power. One can perhaps say that the radiated energy is transported all the way to infinity. On the other hand, if the surrounding medium is lossy, the radiated power is exponentially decaying, and it is sometime assumed that the usual definition of the radiation resistance does not apply, because at the infinite distance from the antenna  the radiated fields are zero.  One can perhaps say that all the power is dissipated in the antenna vicinity. Within this interpretation, we have a confusing ``discontinuity'': If the medium is lossless, \emph{all} the radiated power is transported to infinity, but if the loss factor is nonzero (even arbitrarily small),  \emph{no} power is transported to infinity.

Although there is extensive classical literature on antennas in absorbing media (see the monograph \cite{King} and e.g.~Refs.~\cite{RKMoore,waitieee,karlssonieee,collin,king2}), we have found only limited and sometimes even contradictory discussions on the definition and physical meaning of key parameters of antennas in absorbing media, such as the input impedance and radiation resistance~\cite{RKMoore,Amesieee,Tsao,Des,CA,EC}. In this tutorial paper we carefully examine the notion of radiation resistance of a Hertzian dipole in infinite isotropic homogeneous media and discuss the physical meaning of this model in both lossless and lossy cases. We explicitly calculate the absorbed power in the infinite space and show that the radiation resistance as a model of power transported to infinity is also nonzero for dipoles in lossy media. Subsequently, we study the limit of zero loss. Finally, we discuss implications of this theory for understanding and engineering power transfer between antennas in media.

From the applications point of view, antennas embedded in lossy media have received attention in geophysics, marine technology~\cite{king1,king2,wheelerieee}, medical engineering~\cite{samiieee,nikolayevieee,skrivervik}, etc. Hence, it is worthy to develop a thorough understanding of radiation from the Hertzian dipole which is placed in a lossy medium. These results will also help to understand sub-wavelength emitters or nanoantennas immersed in dissipative media, since we can make an analogy between the Hertzian dipole and the sub-wavelength emitter/nanoantenna that operates in the optical range. The investigation of light interaction with plasmonic or all-dielectric nanoantennas is mainly conducted in the  assumption that the host medium is not dissipative~\cite{maier,bohren,novotny}. However, from the practical point of view, in most applications nanoantennas (nanoparticles) are placed in absorptive environments (e.g. see Refs.~\cite{miller,tanner}). This review is also relevant to the studies of absorption and scattering by small particles in lossy background \cite{mundy,chylek,bohrengilra,sudiarta,fu,yang,svennordebo,svennordebojap}.

If a point source dipole is embedded in a lossy host, there is also a theoretical problem of singularity of power absorbed in the medium, which is due to singularity of the dipole fields at the source point. This issue has been discussed e.g. in \cite{collin}.  The problem of source singularity should be addressed taking into account the final size and shape of the antenna. 
Here, we focus our discussion on radiation phenomena, studying power absorbed outside of a small sphere centred at the source point.


\section{Radiation from Hertzian dipoles in isotropic homogeneous lossy media}

\begin{figure}[t!]\centering
\includegraphics[width=0.4\textwidth]{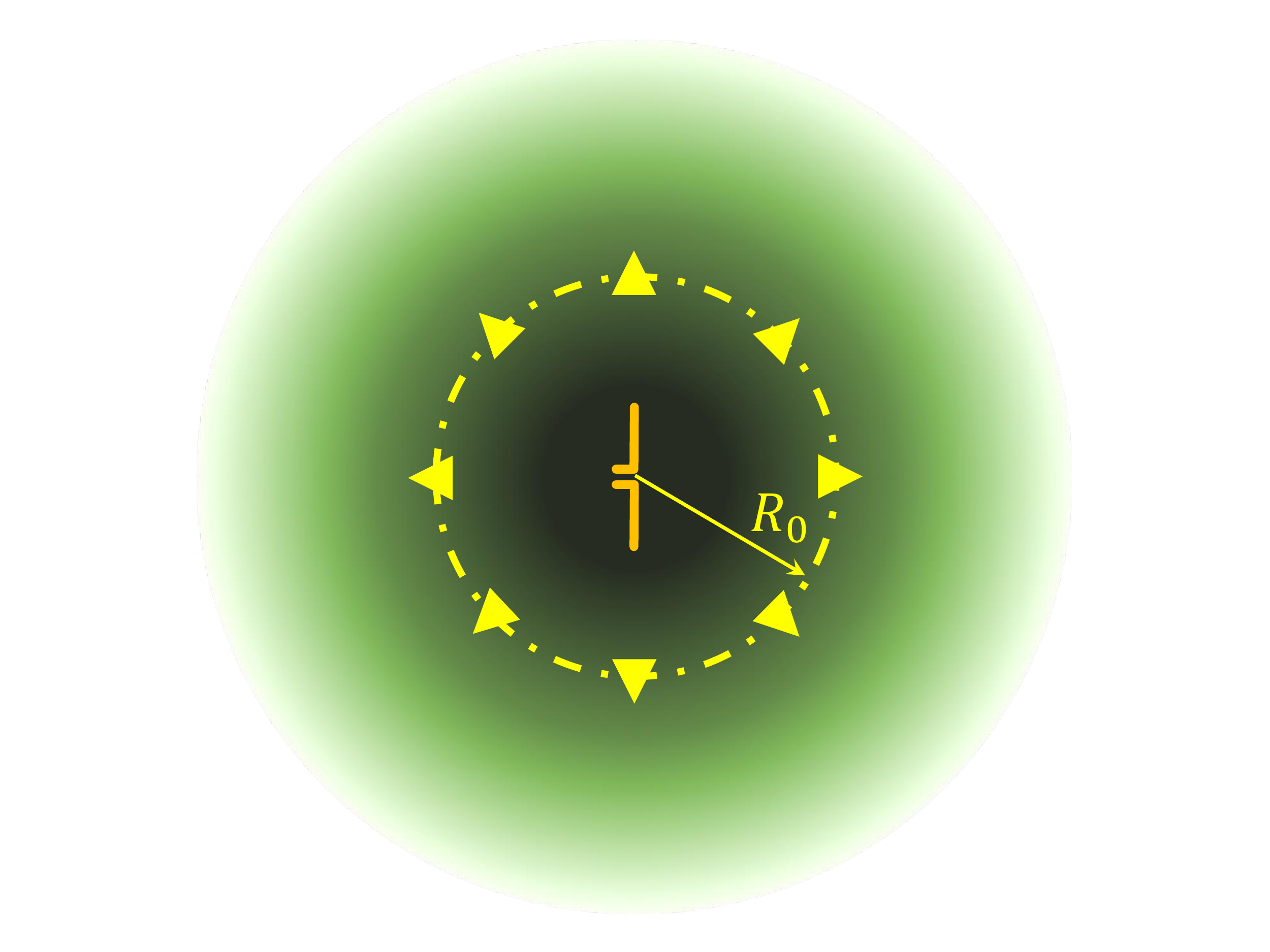}
\caption{The Hertzian dipole is radiating in a lossy medium.}
\label{fig:theory powerdipole}
\end{figure}

Let us consider a Hertzian dipole antenna located in an infinite homogeneous space, as illustrated in Fig.~\ref{fig:theory powerdipole}. The current amplitude in the dipole is fixed and denoted by $I$ (we consider the time-harmonic regime, assuming $e^{j\omega t}$ time dependence). The dipole length is $l$. For simplicity, we assume that the medium is non-magnetic ($\mu=\mu_0$), which does not compromise the generality of our discussion. We characterize the background medium by its complex relative permittivity $\epsilon=\epsilon'-j\epsilon''=\epsilon'-j\sigma/(\omega\epsilon_0)$, where $\sigma$ is the effective conductivity. We will also use the complex refractive index $n=\sqrt{\epsilon}=n'-jn''$ (the square root branch is defined along the positive real axis so that $n''\ge 0$).

The power delivered from the ideal current source $I$ is usually written in terms of the equivalent resistance $R_{\rm in}$, as 
\e 
P_{\rm rad}={1\over 2}R_{\rm in}|I|^2.
\label{e1}
\f
In fact, the resistance $R_{\rm{in}}$ is the real part of the ratio of the voltage to electric current at the input terminals of the dipole~\cite{balanis}. Hence, $R_{\rm{in}}$ has the meaning of the real part of the dipole \emph{input} impedance. We remind that the Hertzian dipole model assumes that the antenna current is fixed, that is, it is an ideal current source without any dissipation inside the antenna structure. If the infinite medium surrounding the dipole is lossless,  such as free space, the input resistance is identical to the radiation resistance. In this case, the usual calculation of the Poynting vector flux through a fictitious spherical surface (which is easy to evaluate in the far zone, see, e.g.~\cite{balanis}) results in
\begin{equation}
R_{\rm{in}}({\rm{lossless}})=R_{\rm{rad}}=\eta_0{(k_0l)^2\over 6\pi}n',
\label{eq:Riinn0}
\end{equation}
in which $R_{\rm{rad}}$ represents the radiation resistance. Also, $\eta_0$ and $k_0$ are the free-space intrinsic impedance and wave number, respectively. Note that the refraction index $n=n'$ is a real, non-negative number\footnote{In isotropic non-magnetic media characterized by relative permittivity $\epsilon$, the choice of the square root branch for $n=\sqrt{\epsilon}$ according to the passivity condition $n''\ge 0$ ensures that $n'\ge 0$. Thus, the resistance (\ref{eq:Riinn0}) is always non-negative. The real part of the refraction index $n'$ is negative in double-negative media, where the real parts of both permittivity and permeability are negative. However, also in that case the resistance value in (\ref{eq:Riinn0}) is non-negative, because for lossless media we get $R_{\rm{in}}=\eta_0{(k_0l)^2\over 6\pi}\mu n$, where both $\mu$ and $n$  are negative.}. However, if the medium is dissipative, the situation may remarkably change. The input resistance $R_{\rm{in}}$ may differ extremely from the radiation resistance since it must also model the dissipation in the near zone. In the following, this difference will be discussed in detail.

\subsection{Time-averaged power and energy conservation}
 
If the host medium is lossy and absorbs power, the calculated ``radiated'' power depends on the radius of the fictitious spherical surface,  and careful considerations are necessary. Since the fields of the point source are singular at the source location, we consider the outward power flux through the surface of a small sphere of the radius  $R_0$, with the dipole at its center, as shown in Fig.~\ref{fig:theory powerdipole} using yellow color. The law of energy conservation tells that all the outward flowing power must be absorbed by the exterior infinite volume ($r>R_0$ in Fig.~\ref{fig:theory powerdipole}). In other words,
\begin{equation}
P_{\rm{rad}}=P_{\rm{abs}},
\label{pt}
\end{equation}
where 
\begin{equation}
P_{\rm{rad}}={1\over2}\oint_{S_0}{\rm{Re}}\big[\mathbf{E}\times\mathbf{H}^*\big]\cdot \mrm{d}\mathbf{S}
\label{eq:eqpovec}
\end{equation}
and 
\begin{equation}
P_{\rm{abs}}={1\over2}\int_V\sigma\vert\mathbf{E}\vert^2\mrm{d}V.
\label{eq:eqpabsabs}
\end{equation}
Here, $S_0$ denotes the surface of radius $R_0$ and $V$ represents the volume of the spherical shell extending from $r=R_0$ to $r\rightarrow\infty$. Furthermore, $\sigma$ is the finite conductivity of the lossy medium. Since Eq.~(\ref{pt}) must always hold for a lossy medium, it should be also true in the limiting case when the dipole is located in a lossless medium. It may be difficult to perceive because the conductivity would be zero ($\sigma=0$) in this case and the right-hand side of Eq.~(\ref{pt}) seems to vanish ($P_{\rm{abs}}=0$). However,  the left-hand side is apparently not zero ($P_{\rm{rad}}\neq0$). Thus, care should be taken in considering the limit of zero losses. Let us first evaluate the power radiated from the sphere of radius $R_0$, given by Eq.~(\ref{eq:eqpovec}) for the general case of lossy media when $\sigma\neq 0$, and then study the case when $\sigma$ tends to zero. This surface integral was calculated in Ref.~\cite{collin}. 
If we substitute the known expressions of the electric and magnetic fields~\cite{balanis}
\begin{equation}
\begin{split}
&E_{r}=-j{Il\over2\pi\omega\epsilon_0\epsilon}\cos\theta\left({1\over r^3}+j{k\over r^2}\right)\exp({-jkr}),\cr
&E_{\theta}=-j{Il\over4\pi\omega\epsilon_0\epsilon}\sin\theta\left({1\over r^3}+j{k\over r^2}-{k^2\over r}\right)\exp({-jkr}),\cr
&H_{\phi}={Il\over4\pi}\sin\theta\left({1\over r^2}+j{k\over r}\right)\exp({-jkr}),\cr
\end{split}
\end{equation}
generated by the Hertzian dipole into Eq.~(\ref{eq:eqpovec}) considering that $k=k_0n$, and calculate the integral, the result reads
\begin{equation}\label{eq:powerpoynting}
\begin{split}
&P_{\rm{rad}}=\cr
&\eta_0{(k_0l)^2|I|^2\over 12\pi}n'\left[{2n''\over (n'^2+n''^2)^2k_0^3R_0^3}+{4n''^2\over (n'^2+n''^2)^2k_0^2R_0^2}\right.\cr
&\left.+{2n''\over (n'^2+n''^2)k_0R_0}+1\right]e^{-2k_0n''R_0}.
\end{split}
\end{equation}
Derivation of the above equation is straightforward (see Supplementary Information). However, calculating $P_{\rm{abs}}$ does not appear to be that simple. Paper \cite{collin} contains a statement that equality (\ref{pt}) holds, but calculation of the volume integral (\ref{eq:eqpabsabs}) is not given, and the limit of $\sigma\rightarrow 0$ is not discussed.

To calculate the integral (\ref{eq:eqpabsabs}), we firstly write the square of the absolute value of the electric field components, which gives 
\begin{equation}
\begin{split}
&\vert E_r\vert^2={|I|^2l^2\cos^2\theta\over 4\pi^2\omega^2\epsilon_0^2(n'^2+n''^2)^2}\cdot\cr&
\bigg({1\over r^6}+{2k_0n''\over r^5}+{k_0^2(n'^2+n''^2)\over r^4}\bigg)e^{-2k_0n''r},
\end{split}
\end{equation}
and 
\begin{equation}
\begin{split}
&\vert E_\theta\vert^2={|I|^2l^2\sin^2\theta\over 16\pi^2\omega^2\epsilon_0^2(n'^2+n''^2)^2}\cdot\cr&
\bigg({1\over r^6}+{2k_0n''\over r^5}+{k_0^2(3n''^2-n'^2)\over r^4}\cr
&+{2k_0^3n''(n'^2+n''^2)\over r^3}+{k_0^4(n'^2+n''^2)^2\over r^2}\bigg)e^{-2k_0n''r}.
\end{split}
\end{equation}
Using the above expressions, we find the square of the absolute value of the electric field, and next  we can evaluate the volume integral in Eq.~(\ref{eq:eqpabsabs}) to calculate the power $P_{\rm{abs}}$ absorbed in the exterior infinite spherical shell. Upon integration over the angles, we find that (see Supplementary Information)
\begin{equation}\label{eq:sigma}
P_{\rm{abs}}={ |I|^2l^2 \over 2\pi\omega\epsilon_0}{n'n''\over (n'^2+n''^2)^2}\int_{R_0}^\infty f(r)e^{-2k_0n''r}\mrm{d}r,
\end{equation} 
where $f(r)=f_1(r)+f_2(r)+f_3(r)$, with 
\begin{equation}\label{eq:sumf}
\begin{split}
&f_1(r)={1\over r^4}+{2k_0n''\over r^3}+{k_0^2(n'^2+5n''^2)\over 3r^2},\cr
&f_2(r)={2k_0^3n''(n'^2+n''^2)\over 3r},\cr
&f_3(r)={k_0^4(n'^2+n''^2)^2\over3}.
\end{split}
\end{equation}
Here, we have used the relation between the imaginary part of the permittivity and the corresponding conductivity
\begin{equation}
\epsilon''=2n'n''={\sigma\over\omega\epsilon_0},
\end{equation}
which gives 
\begin{equation}
\sigma=2\omega\epsilon_0n'n''.
\end{equation}
Note that the term $f_3$, which does not depend on the distance, comes from the $1/r $ terms in the expression for the electric field. To simplify Eq.~(\ref{eq:sigma}), we use partial integration, and derive the following identity: 
\begin{equation}
\begin{split}
&\int_{R_0}^\infty{1\over r^{m}}e^{-\kappa r}=\cr
&\sum_{i=1}^{m-1}{(-1)^{i-1}\kappa^{i-1}e^{-\kappa R_0}\over P(m-1,\,i)R_0^{m-i}}+{(-1)^{m-1}\kappa^{m-1}\over (m-1)!}\int_{R_0}^\infty{e^{-\kappa r}\over r}\mrm{d}r,
\end{split}
\label{eq:operatori}
\end{equation}
where $P(x,\,y)$ refers to the permutation formula. The function $f_1(r)$ consists of three terms and for each of them we can use the expression given by Eq.~(\ref{eq:operatori}). Hence, 
\begin{equation}
\begin{split}
1:\,&\int_{R_0}^\infty{e^{-\kappa r}\over r^4}dr=\bigg[{1\over 3R_0^3}-{k_0n''\over 3R_0^2}+{2k_0^2n''^2\over 3R_0}\bigg]e^{-\kappa R_0}\cr
&-{4k_0^3n''^3\over 3}\int_{R_0}^\infty{e^{-\kappa r}\over r}\mrm{d}r,
\end{split}
\label{eq111}
\end{equation}
\begin{equation}
\begin{split}
2:\,&2k_0n''\int_{R_0}^\infty {e^{-\kappa r}\over r^3}\mrm{d}r=\bigg[{k_0n''\over R_0^2}-{2k_0^2n''^2\over R_0}\bigg]e^{-\kappa R_0} \cr
&+{4k_0^3n''^3}\int_{R_0}^\infty {e^{-\kappa r}\over r}\mrm{d}r,
\end{split}
\label{eq222}
\end{equation}
and 
\begin{equation}
\begin{split}
3:\,&{k_0^2(n'^2+5n''^2)\over 3}\int_{R_0}^\infty {e^{-\kappa r}\over r^2}dr={k_0^2(n'^2+5n''^2)\over 3R_0}e^{-\kappa R_0}\cr
&-{2k_0^3n''(n'^2+5n''^2)\over 3}\int_{R_0}^\infty {e^{-\kappa r}\over r}\mrm{d}r,
\end{split}
\label{eq333}
\end{equation}
where we have introduced the notation  $\kappa=2k_0n''$. We can combine the above results (\ref{eq111}--\ref{eq333}) and write 
\begin{equation}
\begin{split}
&\int_{R_0}^\infty f_1(r)e^{-2k_0n''r}\mrm{d}r=\cr
&\bigg[{1\over 3R_0^3}+{2k_0n''\over 3R_0^2}+{k_0^2(n'^2+n''^2)\over 3R_0}\bigg]e^{-\kappa R_0} \cr
&-{2k_0^3n''(n'^2+n''^2)\over 3}\int_{R_0}^\infty {e^{-\kappa r}\over r}\mrm{d}r.
\end{split}
\label{eq:f1}
\end{equation}
Comparing  Eq.~(\ref{eq:f1})  with Eq.~(\ref{eq:sumf}) we find that the last term in Eq.~(\ref{eq:f1}) (the coefficient before the integral) is the same as function $f_2(r)$ but with the opposite sign. Therefore, only the first three terms inside the square brackets in Eq.~(\ref{eq:f1}) remain and, interestingly, the last term vanishes. Thus,
\begin{equation}\label{eq:f1+f2}
\begin{split}
&\int_{R_0}^\infty \big[f_1(r)+f_2(r)\big]e^{-\kappa r}\mrm{d}r=\cr
&\bigg[{1\over 3R_0^3}+{2k_0n''\over 3R_0^2}+{k_0^2(n'^2+n''^2)\over 3R_0}\bigg]e^{-\kappa R_0}.
\end{split}
\end{equation}
The last step in this long derivation of  $P_{\rm{abs}}$ in Eq.~(\ref{eq:sigma}) is the evaluation of the integral for function $f_3(r)$, which  gives 
\begin{equation}\label{eq:intf3}
\int_{R_0}^\infty f_3(r)e^{-\kappa r}\mrm{d}r={k_0^3(n'^2+n''^2)^2\over 6n''}e^{-\kappa R_0}.
\end{equation}
Using Eqs.~(\ref{eq:f1+f2}) and (\ref{eq:intf3}) and the identity  $\eta_0/k_0=1/(\omega\epsilon_0)$, we  finally 
arrive to the following formula for the power absorbed in the exterior environment (in the spherical shell extending from $R_0$ to infinity):
\begin{equation}\label{eq:finalrespower}
\begin{split}
&P_{\rm{abs}}=\cr
&\eta_0{(k_0l)^2|I|^2\over 12\pi}\bigg[{2n'n''\over (n'^2+n''^2)^2(k_0R_0)^3}\cr
&+{4n'n''^2\over (n'^2+n''^2)^2(k_0R_0)^2}+{2n'n''\over (n'^2+n''^2)(k_0R_0)}+n'\bigg]e^{-\kappa R_0}.
\end{split}
\end{equation}
Now, if we compare Eq.~(\ref{eq:finalrespower}) with Eq.~(\ref{eq:powerpoynting}), we see that they are identical, meaning that Eq.~(\ref{pt}) holds for any non-zero value of the conductivity (or the imaginary part of the permittivity), including the limiting case as $\sigma$ tends to zero. 

Considering the expression (\ref{eq:finalrespower}) for the absorbed power, we note that there are several terms which are proportional to $n''$ and singular at $R_0\rightarrow 0$. They can be interpreted as the power absorbed in the near vicinity of the dipole. The singularity is due to the fact that the fields of a point dipole are singular at the dipole position. Thus, if the medium is lossy ($n''\neq 0$), the absorbed power diverges. Naturally, these terms tend to zero if $n''\rightarrow 0$.

\subsection{Radiation resistance}  

The last term in Eq.~(\ref{eq:finalrespower}), $\eta_0{(k_0l)^2|I|^2\over 12\pi}n'e^{-2k_0n''R_0}$, depends on $n''$ and $R_0$ only in the exponential factor, in contrast to the singular near-field terms. The exponential factor tends to unity when either $R_0$ or $n''$ tends to zero. 
Because this last term is not singular, we can 
let $R_0\rightarrow0$ and interpret this term as the dipole radiation resistance, which measures the ``power delivered to infinity''. Therefore, 
\begin{equation}
R_{\rm{rad}}=2{\lim_{R_0\rightarrow0}\Big[\displaystyle\eta_0{(k_0l)^2|I|^2\over12\pi}n'e^{-2k_0n''R_0}\Big]\over|I|^2}=\eta_0{(k_0l)^2\over6\pi}n'.
\end{equation}  
Importantly, this term results from integration of $1/r$ (``wave'') terms in the expression for the dipole fields, and it is exactly the same as the power radiated into infinite lossless media (see Eq.~\eqref{eq:Riinn0}). Thus, the radiation resistance of electric dipoles in lossy media does not depend on the imaginary part of the refractive index, and it is  expressed by the same formula as for lossless background. This formula for the radiation resistance was given in paper~\cite{Tsao} (the first term in Eq.~(10)), derived on the basis of re-normalizing the power flow density, which is not offering clear physical insight. It also agrees with the result presented in paper~\cite{Des}, Eq.~(2)\footnote{Note the misprint in Eq.~(2) of \cite{Des}: the right-hand side should read $Z(n\omega,\epsilon_0)/n$.}, which is basically equivalent to substituting complex-valued material parameters into the formula derived for the free-space background. On the other hand, contradictory expressions can be also found in the literature, for example, in Refs.~\cite{CA,EC}.

\section{Discussion}

As we see from Eq.~\eqref{eq:finalrespower}, the absorbed power calculated as the volume integral 
\eqref{eq:eqpabsabs} is not zero even in the limit of zero conductivity (lossless background). This result seemingly contradicts to the fact that for $\sigma=0$ the integrand of \eqref{eq:eqpabsabs} is identically zero. However, we cannot conclude that in this case the integral is zero, because for lossless background the integral $\int_V\vert\mathbf{E}\vert^2dV$ diverges. Care should be taken in considering the limit of vanishing absorption if the absorption volume is infinite. 

Let us discuss this limit. The expression for the power absorbed in the medium has a form of a product of conductivity, which tends to zero in the lossless limit, and the integral of the energy density, which diverges when we extend the integration volume to infinity. Thus, we should write the power  balance relation considering absorption in a sphere of a finite radius $R$ and then properly calculate the limit. For this consideration, the energy conservation relation \eqref{pt} is written as 
\begin{multline}\label{eq:Pabsinf}
P_{\rm rad}=\frac{1}{2}\int_{V_R\setminus V_{R_0}}\sigma\left| {\mathbf E} \right|^2\mrm{d}V\\
+\frac{1}{2}\oint_{S_R}\Re\left\{\mathbf{E}\times\mathbf{H}^*\right\}\cdot\mrm{d}\mathbf{S},
\end{multline}
where $\sigma=\omega\epsilon_0\epsilon''=2\omega \epsilon_0n'n''$, and $V_R$ and $V_{R_0}$ denote spheres of radii $R$ and $R_0$, respectively. 
The first term on the right-hand side of \eqref{eq:Pabsinf} is the power dissipated in the surrounding medium between spherical radii $R_0$ and $R$,
and the second term is the power radiated away beyond radius $R$. 
Now we can accurately calculate the limit for $R\rightarrow \infty$ and vanishing $\sigma$ (and $n''$). 
Calculating the integral \eqref{eq:sigma} with a finite upper limit $R$ and considering vanishing $n''$ and infinitely growing $R$, the power balance relation \eqref{eq:Pabsinf} takes the form
\begin{multline}\label{eq:Pabsinf_lim}
P_{\rm rad}=\eta_0{(k_0l)^2 |I|^2\over 12\pi} n'\left(1-\lim_{n''\rightarrow 0,\ R\rightarrow \infty}e^{-2k_0n''R}\right)\\
+\eta_0{(k_0l)^2 |I|^2\over 12\pi} n'\left(\lim_{n''\rightarrow 0,\ R\rightarrow \infty}e^{-2k_0n''R}\right).
\end{multline}
Here, the first line gives the power absorbed in the infinite space, and the second line is the power crossing the spherical surface of the infinite radius.
Clearly, the value of the double-limit depends on the order in which the two limits are taken. Thus, there are two possible interpretations of the absorption in the lossless infinite space. 

1. Taking first the limit $R\rightarrow \infty$ and then the limit of vanishing loss factor  $n''\rightarrow 0$, in the spirit of the principle of vanishing absorption \cite{Vainberg}, we can say that in  this interpretation all the radiated power is dissipated in the vacuum and there is no radiation to infinity (because the expression in the second line tends to zero).

2. Taking first the limit $n''\rightarrow 0$ and then  extending the volume integration over the whole space letting $R\rightarrow \infty$, the expression in the first line (absorption in the medium) tends to zero.  This is a very common interpretation of the lossless infinite space. There are no losses in the ``vacuum'', and all the radiated power is radiated away throughout the infinite space, beyond any finite radius $R$ (in this sense ``transported all the way to infinity'').

However, we stress that in the expression for the  total ``absorbed power'' \eqref{eq:Pabsinf_lim} the two double-limit expressions cancel out, meaning that whatever is the interpretation, the radiation resistance defined as $2P_{\rm rad}/|I|^2$ is a well-behaving, continuous function of $\sigma$ or $n'' $, and based on the first interpretation above we can use the expression \eqref{eq:finalrespower} for both lossy and lossless backgrounds\footnote{Actually, there can be infinitely many intermediate ``intepretations'' since this double-limit expression can take any value from zero to unity depending on the way of taking the limit.}. 

Another, perhaps more important, implication of this consideration is that the input resistance given by Eq.~(\ref{eq:Riinn0}) for \emph{lossy} background contains exactly the same term as the radiation resistance of Hertzian dipole antenna in lossless media or vacuum. Thus, it is misleading to assume that if the background medium is lossy, there is no radiation resistance as such, as the fields exponentially decay. We see that the total input resistance is the sum of the radiation resistance, proportional to $n'$, and the near- and intermediate-zone loss resistance, proportional to $n''$. Importantly, as already discussed, the term proportional to $n'$ does not vanish if $n''$ becomes zero, and it has exactly the same form in both lossy and lossless cases. 

The only scenario where there is no radiation resistance at all is when the \emph{real part} of the refraction index $n'$ vanishes.  Perhaps counter-intuitively, this case corresponds to \emph{lossless} background media. Indeed, calculating 
\e \epsilon=n^2=n'^2-n''^2-2jn'n'',\f
we see that if $n'=0$, then $\epsilon=-n''^2$ is purely real, meaning that the medium is lossless. We also note that the permittivity is negative that is the case when wave propagation is not possible.

Formula \eqref{eq:finalrespower} can be written in terms of the contribution to the real part of the input impedance of the antenna due to dissipation in the surrounding medium (upon dividing by $|I|^2/2$):
\begin{equation}\label{eq:finalrespower_R}
\begin{split}
&R_{\rm in}=\cr
&\eta_0{(k_0l)^2\over 6\pi}n'+\eta_0{(k_0l)^2\over 6\pi}n'n''\bigg[{2\over (n'^2+n''^2)^2(k_0R_0)^3}+\cr
&{4n''\over (n'^2+n''^2)^2(k_0R_0)^2}+{2\over (n'^2+n''^2)(k_0R_0)}\bigg].
\end{split}
\end{equation}
Here, we have assumed that $k_0R_0$ is very small so that $e^{-\kappa R_0}\approx 1$. As mentioned before, the first term is the same as the radiation resistance in lossless media, and can be interpreted as the radiation resistance in \emph{lossy} media. The second term, singular at $R_0\rightarrow 0$, is due to dissipation in the near and intermediate zones. This second term naturally vanishes if $n''=0$. Thus, 
\begin{equation}
R_{\rm{in}}=R_{\rm{rad}}+R_{\rm{nrad}},
\end{equation}
where the non-radiative resistance is given by 
\begin{equation}
\begin{split}
R_{\rm{nrad}}&=\eta_0{(k_0l)^2\over6\pi}n'n''
\bigg[{2\over(n'^2+n''^2)^2(k_0R_0)^3}+\cr
&{4n''\over(n'^2+n''^2)^2(k_0R_0)^2}+{2\over(n'^2+n''^2)(k_0R_0)}\bigg].
\end{split}
\end{equation} 
This consideration clarifies the physical meaning of radiation resistance in the general case of isotropic background media: It vanishes only if \emph{propagation is not possible}. In other words, the radiation resistance is not zero even in lossy media, because the propagation constant is not equal to  zero and the outward power flux is not zero. 

In terms of applications, this conclusion is important for understanding of ultimate limits for power transported from one antenna to another. Let us position a receiving Hertzian dipole antenna of length $l$ in  the field created by our radiating dipole antenna (in an infinite isotropic medium characterized by the refractive index $n=n'-jn''$). The current induced in the receiving antenna $I_{\rm rec}$ is proportional to the electric field $E$ created by the transmitting antenna at the receiver position and inversely proportional to the impedance:
\e I_{\rm rec}={El\over R_{\rm load}+ jX+R_{\rm in}}.\f
Let us assume that the receiving antenna is loaded by a resistor $R_{\rm{load}}$ at its center. The power delivered to the load reads 
\e P={1\over 2}R_{\rm{load}}|I_{\rm rec}|^2.\f
Obviously, to maximize the delivered power we should bring the antenna to resonance, making the total reactance $X=0$, which is always possible. The ideal scenario where the delivered power can be arbitrarily high corresponds to $R_{\rm in}=0$ (then we have $P=|E|^2l^2/R_{\rm{load}}$ which diverges for $R_{\rm{load}}\rightarrow 0$). Now we should look at the expression for the effective input resistance and find under what conditions this expression gives the smallest value. Apparently, conventional lossless background ($n''=0$) is better than the lossy one, but even in that case the received power is limited, because the radiation resistance (\ref{eq:Riinn0}) is not zero. This consideration brings us to the well-known limit of the effective absorption cross section for any dipole antenna/scatterer  in lossless background \cite{sergei}. But notice an important special case of $n'=0$. In this case the input resistance is zero, and the delivered power has no upper bound. 

Accordingly, we reach an enlightening conclusion. If two antennas are in an environment which  \emph{does not} allow wave propagation, the power delivered from one antenna to the other can be arbitrarily high. In contrast, if wave  propagation is allowed, the delivered power is fundamentally limited by the ultimate absorption cross section of a dipole antenna, which, in turn, is determined by its non-zero radiation/input resistance. Another case when the power delivered to a load from an ideal current source is limited only by the parasitic resistance of conductors, is the case of zero frequency (DC) circuits. Interestingly, the reason why there is no fundamental limit is the same: There is no radiation loss at DC.

The scenario of such unlimited-capacity power delivery channel corresponds, for instance, to the case of two dipole antennas inside a lossless-wall waveguide below cut-off. The two antennas are coupled only by reactive fields, and at the resonance of the antenna pair inside the waveguide, all the effective reactances are compensated. At this frequency, the receiving antenna is effectively connected to the ideal current source feeding the transmitting antenna by a zero-impedance link. Here, the power delivered from one antenna to the other is limited only by the power available from the source, and by parasitic losses in the waveguide walls and in the antenna wires. Another example is coupling between two antennas in lossless plasma. 

\section{Conclusions}

In this tutorial review we have addressed the question of radiation resistance of electric dipole antennas in lossy background. We have shown that the power absorbed in the background medium (excluding a small sphere of radius $R_0$ around the source) contains terms which diverge at $R_0\rightarrow 0$ and measure the power delivered by the antenna near fields, and, importantly, the  radiation resistance term, which measures the power delivered to the lossy host by the wave fields (decaying as $1/r$). An important conclusion is that this radiation resistance does not depend on the imaginary part of the refractive index: It differs from the well-known formula for the radiation resistance of a dipole in free space by multiplication by the real part of the refractive index. 
We have discussed in detail the limit of zero loss factor, comparing two possible interpretation of power ``loss'' in lossless vacuum. 
Finally, we presented some considerations on fundamental limitations on power coupling between two antennas and possible scenarios of removing the limit imposed by radiation damping.


\end{document}